\documentclass[aip,apl,a4paper,reprint,footinbib]{revtex4-1}

\usepackage[utf8]{inputenc}
\usepackage[T1]{fontenc}
\usepackage{amssymb}
\usepackage{amsmath}
\usepackage{amsbsy}
\usepackage{bm}
\usepackage{graphicx}

\usepackage{docs}
\usepackage[colorlinks=true,linkcolor=red]{hyperref}%
\expandafter\ifx\csname package@font\endcsname\relax\else
 \expandafter\expandafter
 \expandafter\usepackage
 \expandafter\expandafter
 \expandafter{\csname package@font\endcsname}%
\fi
\newcommand{\pe}{\! = \!}

\newcommand{\vn}{\bm{n}}
\newcommand{\vz}{\bm{0}}

\newcommand{\vk}{\bm{k}}

\newcommand{\mk}{|\vk|}

\newcommand{\cH}{\mathcal{H}}
\newcommand{\gt}{\tilde{\gamma}}

\newcommand{\hC}{h_{{\scriptscriptstyle{\textrm{C}}}}}
\newcommand{\td}{t_{{\scriptscriptstyle{\textrm{d}}}}}

\newcommand{\hD}{h_{{\scriptscriptstyle{\textrm{D}}}}}

\begin{document}

\title{Orientation dependence of the elastic instability on strained SiGe films}

\author{J.-N. Aqua}
\email{jean-noel.aqua@im2np.fr}
\affiliation{Institut Mat\'eriaux Micro\'electronique Nanoscience de Provence, 
 Aix-Marseille Universit\'e,
UMR CNRS 6242, 13997 Marseille, France}
\author{A. Gouyé}
\affiliation{Institut Mat\'eriaux Micro\'electronique Nanoscience de Provence, 
 Aix-Marseille Universit\'e,
UMR CNRS 6242, 13997 Marseille, France}
\author{T. Auphan}
\affiliation{Institut Mat\'eriaux Micro\'electronique Nanoscience de Provence, 
 Aix-Marseille Universit\'e,
UMR CNRS 6242, 13997 Marseille, France}
\author{I. Berbezier}
\affiliation{Institut Mat\'eriaux Micro\'electronique Nanoscience de Provence, 
 Aix-Marseille Universit\'e,
UMR CNRS 6242, 13997 Marseille, France}
\author{T. Frisch}
\affiliation{Institut Mat\'eriaux Micro\'electronique Nanoscience de Provence, 
 Aix-Marseille Universit\'e,
UMR CNRS 6242, 13997 Marseille, France}
\author{A. Ronda}
\affiliation{Institut Mat\'eriaux Micro\'electronique Nanoscience de Provence, 
 Aix-Marseille Universit\'e,
UMR CNRS 6242, 13997 Marseille, France}

\date{\today}
\begin{abstract}
At low strain, SiGe films on Si substrates undergo 
a continuous nucleationless morphological evolution known as the Asaro-Tiller-Grinfeld instability. 
We demonstrate experimentally that this instability develops on Si(001) 
but not on Si(111) even after long annealing. Using a continuum description of this instability, we determine the origin of this difference. When modeling surface diffusion in presence of wetting, elasticity and surface energy anisotropy, we find a retardation of the instability on Si(111) due to a strong dependence of the instability onset as function of the surface stiffness. This retardation is at the origin of the inhibition of the instability on experimental time scales even after long annealing.
\end{abstract}

\maketitle


The complexity of electronic devices has grown extensively during the last decades.
The fundamental limitations for further device improvement challenge the physical
mechanisms at stake in the nanoscales. In CMOS technology, 
silicon-germanium (SiGe)-based materials have shown to overwhelm silicon-only materials and 
to compete seriously with the 
high performances of III-V compounds (cut-off frequencies, current gain...). 
The understanding of the morphological evolution of thin SiGe films 
thus displays a particular interest for potential application.

The coherent epitaxy of Si$_{1-x}$Ge$_x$ films on Si substrates follows different paths 
depending on the growth parameters:\cite{StanHoly04,BerbRond09}
a nucleationless morphological instability at low $x$ followed by island formation on top of a wetting layer, 
2D/3D island nucleation at large $x$ after the completion of a thin wetting layer, 
step bunching on vicinal substrates, or plastic relaxation via nucleation of dislocations above a critical height $\hD$. 
We are interested here by the understanding of the elastic instability onset on nominal 
substrates\cite{SuttLaga00,*TromRoss00} and of 
its basic ingredients. The latter is reminiscent of the Asaro-Tiller-Grinfel'd (ATG) instability
first explained in Refs.~\onlinecite{AsarTill72,Grin86,Srol89} by an enhanced surface diffusion driven by 
a combination of elastic relaxation and surface energy reduction.
In addition, for thin films, wetting interactions also come into play and stabilize a 2D wetting film so that the instability can not occur below a thermodynamic critical height $h_c$.\cite{ChiuGao95}
Moreover, it was shown that \textit{as-grown} films on different orientations undergo different 
evolutions:\cite{BerbGall98} while the instability is manifest on Si(001), it is inhibited on Si(111) 
where longer growth eventually leads to dislocations. We investigate here the influence of surface energy anisotropy 
on the dynamics of the instability through annealing experiments that we confront to the continuous theory. 
Considering the extra ingredient of the instability modelling which is the film surface energy anisotropy,  
we argue that the difference between the two orientations 
is attributed to a retardation of the instability on Si(111). The latter is characterized by a strong dependence of 
the instability onset as function of the surface stiffness.

	
Growth experiments were performed by Molecular Beam Epitaxy (MBE) in a Riber system with a base pressure 
of $\sim$10-11\,Torr. After 950$^o$C in situ flashing, a Si buffer layer 50\,nm thick is 
deposited at 750$^o$C to provide a perfectly clean reproducible flat surface. Si flux is obtained from an 
electron beam evaporator and maintained constant during the growth at $\sim$0.1\,ML/s. Ge is deposited from an effusion cell. The growth temperature is 550$^o$C and 
the Si$_{0.85}$Ge$_{0.15}$ composition corresponds to a low misfit ($\sim\!0.6$\%) where 
the elastic instability occurs during growth without onset of island formation. Silicon substrates 
are rotated during the experiments and their temperature is real time recorded. Morphologies $z\pe h(x,y)$ are analyzed 
after growth by atomic force microscopy (AFM) in air in contact and no contact modes. These evolutions were investigated 
on Si(001) and Si(111) just after growth and after in situ thermal annealing at 550$^o$C (up to 20\,h). In the present work 
the thickness (130\,nm) and composition ($x \pe 0.15$) 
of the SiGe layers were adjusted to maintain the structures below the critical 
thickness of dislocation while providing enough strain energy to develop the instability.

\begin{figure}[h] \centering
\includegraphics[width=6.5cm]{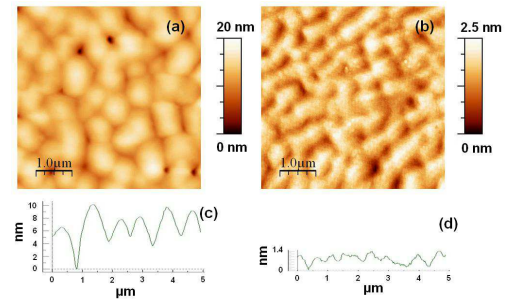}
\caption{(color online) 
AFM top-view images of 130 nm-thick as-grown Si$_{0.85}$Ge$_{0.15}$ 
layers (a) on (001) Si substrate and (b) on (111) Si substrate; side views are shaped in (c) and (d) respectively.}
\label{fig-asgrown}
\end{figure}

\begin{figure}[h] \centering
\includegraphics[height=3.5cm]{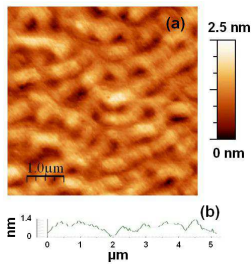}
\caption{(color online) 
 (a) AFM top-view image of 130nm-thick Si$_{0.85}$Ge$_{0.15}$ 
 layers on (111) Si substrate after 20\,h anneal; Side view is shaped in (b).}
\label{fig-anneal}
\end{figure}

The surface of  130\,nm thick Si$_{1-x}$Ge$_x$ films just after growth on Si(111) and Si(100)
are displayed in Fig.~\ref{fig-asgrown}. On Si(001), the elastic instability is already fully developed at the end 
of the 1h long growth, and produces a periodic corrugation characterized by a root mean square
roughness $w \pe \langle (h-\langle h\rangle)^2\rangle^{1/2}$ equal to $w\pe1.87$\,nm. On the contrary, the 
experiment performed in the same experimental conditions on a Si(111) substrate leads
to an almost flat surface with an intrinsic noise characterized by $w \pe 0.33$\,nm.
To investigate the influence of kinetics on the onset of the instability, 20\,h long annealing of the 
2D film deposited on Si(111) was performed which result is displayed in Fig.~\ref{fig-anneal}. 
Annealing did not allow for the development of the instability, since the resulting film is
not significantly changed and is characterized by a roughness 
with $w \pe0.32$\,nm.

To give some insights into the reported difference, we turn to 
the modelling of the ATG instability which is driven by the elastic relaxation enforced by 
a surface corrugation. Its description involves a continuum representation of surface mass 
currents,\cite{AsarTill72,Grin86} where the film height $h(x,y,t)$ evolves due to gradients 
in the chemical potential. The latter includes a surface energy $\gamma (h,\vn)$
which dependences on $h$ and on the local normal to the surface $\vn$ reflect respectively 
wetting interactions and crystalline anisotropy.\cite{jnFris10}
For simplicity, we consider $\gamma (h,\vn) \pe \gamma_f \left[1+\gamma_h (h) +\gamma_{\vn} (\vn)\right]$ 
where $\gamma_f$ is a reference surface energy. 
The elastic chemical potential may be computed in the small-slope approximation 
in terms of the system Green function and of the difference between the film and substrate 
lattice parameters $a^{f/s}$.\cite{jnFris10} 
One can then define the length scale  $l_0 \pe \gamma_f (1-\nu)/ [2 (1+\nu) E (1-a^f/a^s)^2]$ and 
time scale $t_0 \pe l_0^4/(D \gamma_f)$, where $\nu$ and $E$ are the film Poisson ratio 
and Young modulus, and $D$, the surface diffusion coefficient. 
The evolution equation at first order in $h$ is then
\begin{equation}
\label{eqevolanis}
\frac{\partial h}{\partial t} =
 - \boldsymbol{\Delta} \left\{ \rule{0mm}{5mm} 
     \gamma(\vn,h) \bm{\Delta} h
    +h_{ij} \frac{\partial^2 \gamma}{\partial n_i n_j}
    - \frac{\partial\gamma}{\partial h}
    + \cH_{ii} (h)
    \right\} ,
\end{equation}
with $i,j\pe x,y$ and 
where the first and second term in the bracket describe the anisotropic surface energy, 
and the third term, the wetting interactions. The elastic energy density is given by 
the Hilbert transform $\cH_{ii}$ which acts in Fourier space as a multiplication 
by $\mk$.\cite{jnFris10} 

The solution for $h$, with average $h_0$, is given in Fourier space by
$\hat{h} (\vk,t) \pe \hat{h} (\vk, 0) \exp(\sigma \, t)$, with the growth rate  
\begin{equation}
	\label{sigma2k}	
	\sigma(\vk; \vn_0, h_0) = - \mbox{$\frac{\partial^2 \gamma}{\partial h^2}$} (h_0) \, \vk^2 + \mk^3 
					- \gt(\vn_0)  \, \vk^4 ,
\end{equation} 
where the first and last term describe the stabilizing wetting interactions 
and surface energy, and the second term, the long-range destabilizing elastic interactions. 
Anisotropy is embedded in the relative stiffness 
 $\gt (\vn) = \left[ \gamma + \partial^2 \gamma / \partial h_x^2 
	+ \partial^2 \gamma / \partial h_y^2 \right]/\gamma_f$, 
which is computed in \eqref{sigma2k} at $\vn_0$, the substrate orientation.
\footnote{Note that a small dependance of $\gt$ on $h_0$ should be present but was
discarded as it has little influence on the forthcoming results.} 
Finally, we fit atomistic calculations\cite{LuLiu05} with
$\gamma_h (h) \pe c_w \exp(-h/a^f)$, where, the variation for Ge on Si gives the extrapolation $c_w\pe 0.05$
for $x\pe0.15$. In order to understand the dramatic difference of the film evolution between (001) and (111)
orientations, we focus on the influence of the surface stiffness on the 
instability onset, which is known to differ markedly between these orientations. 

\begin{figure}[h] \centering
\includegraphics[width=7cm,height=3.6cm]{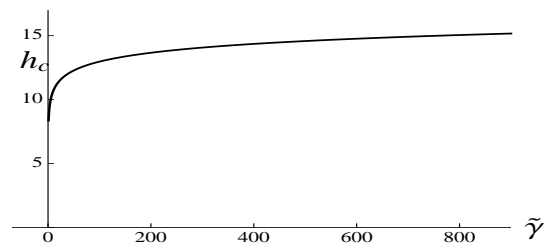}
\caption{Critical height (in ML) above which the elastic instability may develop 
as function of the surface stiffness $\gt$.}
 \label{figstab}
\end{figure}

To study the thermodynamic stability of the film, we first consider the critical height $\hC$ 
above which the morphological evolution may happen as the energy of the corrugated 
layer becomes lower than the one of a flat layer. 
We define $h^*$ above which $\sigma(\vk)$ given in Eq.~\eqref{sigma2k} displays a local maximum beside 
$\vk \pe \vz$, and which depends logarithmically on $\gt$ due to the exponential dependence of $\gamma_h$. 
We then compute $\hC$ numerically 
by searching for the positiveness of $\sigma$ at this local maximum.\footnote{which is, to a good approximation, equal to $h^*$}The value of $\hC$ as function of the surface stiffness is displayed in Fig.~\ref{figstab} 
and is characterized by a logarithmic behavior with little variation over reasonable values of $\gt$. 
Given the large deposited film thickness (above 200\,ML), we conclude that the inhibition of the elastic instability on 
Si(111) cannot be attributed to an energetic effect linked to an enhanced 
stabilization of the dispersion relation by the surface stiffness. 

Considering this conclusion, 
we now study the dynamics of the instability. 
We look for the time $\td$ where the transition between the flat fully-strained and the corrugated layers experimentally occurs.
We characterize the emergence of the instability by the criterion that the roughness 
is greater than some value $w_s$. Given the 
solution \eqref{sigma2k} for $h$ at linear order, the roughness at time $t$ is given by 
$w^2 \pe \left( 2\pi/L\right)^2 \int d^2\vk \, |\tilde{h}_1(\vk,0)|^2 \exp(2 
	\sigma t)$, 
where $L$ is the system size in the $x$ and $y$ directions. We consider for the surface initial 
condition, a white noise of roughness $w_0$ distributed on $L^2$ modes. 
The criterion for the instability to be observable is then merely 
\begin{equation}
	\label{wpgs}
	 w_s^2/w_0^2 = (2 \pi)^{-2} \int_{|k_x, k_y| < \pi}\! \! \! d^2\vk \, \, e^{2 \, \sigma(\vk)\, t} ,
\end{equation}
which can be numerically computed, leading to the time 
$\td$ for the onset of the instability.  

\begin{figure}[h] \centering
\includegraphics[width=7cm,height=3.5cm]{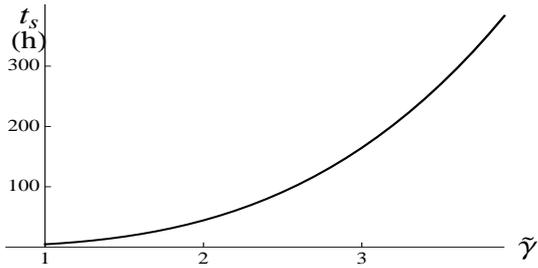}
\caption{Characteristic time for the instability to be fully developed 
as function of the surface stiffness.}
 \label{figretard}
\end{figure}

The resulting $\td$ for a thick film as function of the surface stiffness is plotted in Fig.~\ref{figretard}, 
considering the threshold $w_s/w_0 \pe 5$.
We find that $\td$ increases markedly over a small increase of $\gt$, becoming rapidly beyond 
experimental reachable time scales for relatively modest stiffnesses, revealing a notable retardation 
of the instability dynamics induced by the surface stiffness. 
The inhibition of the elastic instability on Si(111) may hence be attributed to this stiffness-induced 
retardation as the (111) orientation is known to be stiffer than the (001) 
orientation.\cite{Eagl93,BermMeto95,TersSpen02,MoorRetf06}
This higher stiffness of Si(111) is due both to a higher step formation 
energy, between $0.01$\,eV and $0.15$\,eV for the two different step edges on Si(001) and 
$0.19$\,eV on Si(111), and lower step diffusivity (inversely proportional to the step stiffness)\cite{JeonWill99}  
1\AA~on Si(111) and 15\AA~on Si(001). As shown in  
Fig.~\ref{figretard}, the evolution of $\td$ with $\gt$ is so large 
that the system may reach during growth the critical thickness for dislocation nucleation 
before the onset of the morphological evolution. Hence, we attribute the transformation 
during growth on Si(111)\cite{BerbGall98} of 2D layers into dislocated 2D layers to this phenomena.

Our conjecture about the stiffness-induced retardation 
is also supported by the evolution of SiGe layers on vicinal substrates.\cite{BerbRond03}
On vicinal Si(001) substrates the ATG instability follows the same morphological evolution than on 
nominal substrate, but undergoes a morphological change with the transformation of square based ripples into 
1D elongated ripples.\cite{BerbRond03} On the other hand, on vicinal Si(111) (when the angle exceeds 1$^o$),  
the onset of a step bunching instability at low misfits\cite{TeicBean98} which
relaxes strain through edge effects is observed as opposed to the case on nominal substrate reported above. 
Considering the strong effect of the atomic steps on Si(111) which enforce a much lower 
stiffness of the vicinal surface as compared to the nominal one, we advocate that the 
surface stiffness is the essential parameter ruling the dynamics of strained films. 

If a crucial difference between the (001) and (111) orientations concerns their stiffnesses, 
a relevent ingredient which may also 
come into play is the difference in their diffusion coefficients $D$. Indeed, the time scale $t_0$ 
of the instability is inversely proportional to $D$ and a kinetic difference could also originate from 
this coefficient. However, considering the quantitative differences of the diffusion coefficients between 
these two orientations reported in Ref.~\onlinecite{JeonWill99}, the inhibition of the instability on 
Si(111) can not be explained by this effect. Indeed, the time constants for diffusion on Si(001) and Si(111) 
at 950$^o$C are $1.5 \,10^{-4}$s\cite{BartTrom94} and 
$10^{-6}$s\cite{BartGold93} respectively, which testify a higher surface diffusion on 
Si(111) compared to Si(001). 
Finally, other effects such as alloying are unlikely to have a strong enough 
influence on the instability dynamics. 


As a conclusion, we studied the orientation dependence of the morphological ATG instability during MBE of thin SiGe 
strained films in coherent epitaxy on Si. 
While the strain-induced corrugation is fully developed on Si(001) substrates, its counterpart on 
Si(111) appears to be inhibited even after a 20\,h long annealing. We revisited the basic ingredients 
of the elastic instability and concluded that the surface stiffness has a dramatic influence 
as it enforces a significant retardation of the dynamics. 
Considering the time necessary to obtain a large enough roughness, we find that the instability may be 
postponed over experimentally unreachable timescales due to the surface stiffness and we argue that 
this effect causes the absence of the morphological evolution on Si(111).

The authors thank R. Kern for fruitful discussions. 


%

\end{document}